\documentstyle[preprint,eqsecnum,aps]{revtex}
\tightenlines
\newcommand{\bea}{\begin{eqnarray}}
\newcommand{\eea}{\end{eqnarray}}
\newcommand{\beq}{\begin{equation}}
\newcommand{\eeq}{\end{equation}}
\newcommand{\bay}{\begin{array}}
\newcommand{\eay}{\end{array}}
\begin{document}
\preprint{\parbox{6cm}{\flushright CLNS 97/1520\\TECHNION-PH 97-07}}
\title{Excited Baryons Phenomenology from Large-$N_c$ QCD}
\author{Dan Pirjol\footnote{On leave from the Department of Physics, 
Technion - Israel Institute of Technology, 32000 Haifa, Israel} and 
Tung-Mow Yan\footnote{On leave from the Floyd R. Newman Laboratory of Nuclear 
Studies, Cornell University, Ithaca, New York 14853}}
\address{Center for Theoretical Sciences\\
Physics Department\\
National Tsing Hua University, Hsinchu 30043, Taiwan, R.O.C.}
\date{\today}
\maketitle
\begin{abstract}
We present a phenomenological analysis of the strong couplings of the
negative-parity $L=1$ baryons from the perspective of the large-$N_c$
expansion. In the large-$N_c$ limit the mass spectrum and mixing 
pattern of these states are constrained in a very specific way.
The mixing angles are completely determined in this limit, with predictions
in good agreement with experiment. In the combined large-$N_c$ and SU(3) 
limits the pion couplings of the five negative-parity octets to the ground 
state baryons are given in terms of only 3 independent couplings. The 
large-$N_c$ predictions for the ratios of strong couplings are tested 
against experimental data. 
\end{abstract}
\pacs{pacs1,pacs2,pacs3}

\narrowtext
\section{Introduction}

The large-$N_c$ expansion \cite{LargeNc} proved to be a valuable guide for a 
qualitative and even quantitative understanding of gauge theories.
In the past few years its application to baryons in QCD pioneered by
Witten \cite{Wi,Coleman} has been substantiated and greatly expanded in a 
series of papers by Dashen, Jenkins and Manohar (DJM) and others 
\cite{DJM1,DJM2} (and references cited therein).

In a recent paper \cite{1} we studied the strong couplings of the orbitally 
excited baryons in the framework of the large-$N_c$ expansion, extending the 
results obtained by DJM in the s-wave sector. The general structure of 
the pion couplings to these states has been derived from a set of 
consistency conditions which follow 
from requiring the total scattering amplitude to satisfy the Witten scaling 
rules. The analysis presented in \cite{1} assumed only isospin symmetry and 
was for the most part limited to baryons containing only $u$ and $d$ 
quarks. The present paper is a continuation to \cite{1} and its aim is 
two-fold: first, to extend the results of \cite{1} by incorporating
SU(3) symmetry and second, to present a phenomenological analysis of the 
existing experimental data from the perspective of the large-$N_c$ 
expansion. 

In Section II we demonstrate that the combined large-$N_c$ and SU(3) limits of 
QCD provide very strong constraints on the structure of the mass spectrum and 
mixing pattern of the $L=1$ light baryons. 
A set of relations are derived among strong transition amplitudes between 
p-wave and s-wave baryons in Sec.III which are then compared against 
available experimental data. 
These relations are shown explicitly to agree with those derived in the 
quark model with arbitrary number of colors in the limit $N_c\to\infty$.
For $N_c=3$ they reduce to the usual SU(6) predictions of the quark model
\cite{FaiPla,Hey}. However, the large-$N_c$ approach turns out to be both
less and more predictive than the SU(6)-based. On the one hand it predicts 
well-defined values for the mixing angles (which are left completely arbitrary 
in the quark model) but on the other hand, due to the small value of the
number of colors in the real world, its applicability to the decuplet states is 
limited.
One of the large-$N_c$ relations among $S$-wave pion couplings appears to
be badly violated and we discuss a few possible explanations, one of which
involves a different quark model assignment for the observed $S_{11}$ states.
We summarize our conclusions in Sec.IV.

\section{SU(3) spin-flavor structure of the excited baryons}

The structure of the baryon spectrum in the large-$N_c$ limit can
be obtained by examining the symmetry properties of the states 
under permutations of two quarks. 
The ground state s-wave baryons transform according to the completely
symmetric representation of the permutation group shown in Eq.(\ref{SU(6)symm}).
For baryons containing two flavors this means that their spin-flavor 
wavefunction must transform like the totally symmetric representation
of SU(4), which is decomposed into representations of $SU(2)_{isospin}\times
SU(2)_{spin}$ with $I=J$. The analogous decomposition of the totally
symmetric representation of SU(6) into representations of 
$SU(3)_{flavor}\times SU(2)_{spin}$, relevant for the baryons
containing 3 light flavors is shown in Eq.(\ref{SU(6)symm}).
For $N_c=3$ this representation contains the familiar spin-1/2 octet 
and the spin-3/2 decuplet baryons.

% draw box with width #1pt and line thickness #2pt
\newcommand{\drawsquare}[2]{\hbox{%
\rule{#2pt}{#1pt}\hskip-#2pt%  left vertical
\rule{#1pt}{#2pt}\hskip-#1pt%  lower horizontal
\rule[#1pt]{#1pt}{#2pt}}\rule[#1pt]{#2pt}{#2pt}\hskip-#2pt%  upper horizontal
\rule{#2pt}{#1pt}}% right vertical
\bea\label{SU(6)symm}
\overbrace{\,\raisebox{-3.0pt}{\drawsquare{10.0}{0.4}}\hskip-0.4pt
        \raisebox{-3.0pt}{\drawsquare{10.0}{0.4}}\cdots
        \raisebox{-3.0pt}{\drawsquare{10.0}{0.4}}\,}^{N_c} &=&
\left(\,\,
\raisebox{-8.0pt}{\drawsquare{10.0}{0.4}}\hskip-10.4pt
        \raisebox{2pt}{\drawsquare{10.0}{0.4}}\hskip-0.4pt
\raisebox{-8.0pt}{\drawsquare{10.0}{0.4}}\hskip-10.4pt
        \raisebox{2pt}{\drawsquare{10.0}{0.4}}\hskip-0.4pt
\cdots
\raisebox{-8.0pt}{\drawsquare{10.0}{0.4}}\hskip-10.4pt
        \raisebox{2pt}{\drawsquare{10.0}{0.4}}\hskip-0.4pt
        \raisebox{2pt}{\drawsquare{10.0}{0.4}}\hskip-0.4pt
\,,J=\frac12\right) +
\left(\,\,
\raisebox{-8.0pt}{\drawsquare{10.0}{0.4}}\hskip-10.4pt
        \raisebox{2pt}{\drawsquare{10.0}{0.4}}\hskip-0.4pt
\raisebox{-8.0pt}{\drawsquare{10.0}{0.4}}\hskip-10.4pt
        \raisebox{2pt}{\drawsquare{10.0}{0.4}}\hskip-0.4pt
\cdots
\raisebox{-8.0pt}{\drawsquare{10.0}{0.4}}\hskip-10.4pt
        \raisebox{2pt}{\drawsquare{10.0}{0.4}}\hskip-0.4pt
        \raisebox{2pt}{\drawsquare{10.0}{0.4}}\hskip-0.4pt
        \raisebox{2pt}{\drawsquare{10.0}{0.4}}\hskip-0.4pt
        \raisebox{2pt}{\drawsquare{10.0}{0.4}}\hskip-0.4pt
\,,J=\frac32\right)
\eea
\bea\nonumber
\qquad\qquad\qquad +\,\cdots +
\left(\,\,
\raisebox{-3.0pt}{\drawsquare{10.0}{0.4}}\hskip-0.4pt
        \raisebox{-3.0pt}{\drawsquare{10.0}{0.4}}\cdots
        \raisebox{-3.0pt}{\drawsquare{10.0}{0.4}}
\,, J=\frac{N_c}{2}\right)\,.
\eea

The spectrum of the p-wave baryons can be obtained in a similar way from
symmetry considerations. In the real world with $N_c=3$ the spin-flavor 
wavefunction of the $L=1$ light baryons transforms according to the mixed 
symmetry representation {\bf 70} of SU(6). Its decomposition into 
spin-flavor multiplets takes the form \cite{Close}
\bea
\raisebox{-8.0pt}{\drawsquare{10.0}{0.4}}\hskip-10.4pt
        \raisebox{2pt}{\drawsquare{10.0}{0.4}}\hskip-0.4pt
\raisebox{2pt}{\drawsquare{10.0}{0.4}}
 = ({\bf 1}\,, S=\frac12) \oplus ({\bf 10}\,, S=\frac12) \oplus
({\bf 8}\,, S=\frac12) \oplus ({\bf 8}\,, S=\frac32)\,.
\eea
After adding the orbital angular momentum $L=1$ the resulting states 
reproduce the observed spectrum of the p-wave light baryons \cite{PDG}.

We would like in the following to construct the generalization of this 
procedure to the case of arbitrary $N_c$. The corresponding representation
of SU(6) is obtained by adding additional boxes to the first line of the
Young diagram. 
Its decomposition under the flavor-spin SU(3)$\times$SU(2) subgroup 
can be obtained as described in \cite{1} for the corresponding SU(4) 
representation.
One starts with the product of SU(6) representations
\bea\label{SU(6)prod}
\overbrace{\,
\raisebox{-3pt}{\drawsquare{10.0}{0.4}}\hskip-0.4pt
\raisebox{-3pt}{\drawsquare{10.0}{0.4}}\hskip-0.4pt\,\cdots\,
\raisebox{-3pt}{\drawsquare{10.0}{0.4}}\hskip-0.4pt\,}^{N_c-1}
\,\, \otimes\,\,\,
 \raisebox{-3pt}{\drawsquare{10.0}{0.4}}\hskip-0.4pt\,\,\, =\,\,\,
\overbrace{\,
\raisebox{-3pt}{\drawsquare{10.0}{0.4}}\hskip-0.4pt
\raisebox{-3pt}{\drawsquare{10.0}{0.4}}\hskip-0.4pt\,\cdots\,
\raisebox{-3pt}{\drawsquare{10.0}{0.4}}\hskip-0.4pt\,}^{N_c}
\,\, \oplus \,\,\,
\overbrace{\,
\raisebox{-8.0pt}{\drawsquare{10.0}{0.4}}\hskip-10.4pt
        \raisebox{2pt}{\drawsquare{10.0}{0.4}}\hskip-0.4pt
\raisebox{2pt}{\drawsquare{10.0}{0.4}}\,\cdots\,
\raisebox{2pt}{\drawsquare{10.0}{0.4}}\,}^{N_c-1}
\eea
The decomposition of the symmetric representation on the
left-hand side is known from Eq.(\ref{SU(6)symm}). Subtracting from the product 
on the left-hand side the
representations of SU(3)$\times$SU(2) corresponding to the symmetric
representation on the right-hand side we obtain

\bea\label{SU(6)mixed}
\overbrace{\,
\raisebox{-8.0pt}{\drawsquare{10.0}{0.4}}\hskip-10.4pt
        \raisebox{2pt}{\drawsquare{10.0}{0.4}}\hskip-0.4pt
\raisebox{2pt}{\drawsquare{10.0}{0.4}}\,\cdots\,
\raisebox{2pt}{\drawsquare{10.0}{0.4}}\,}^{N_c-1} &=&
\left(\,\, 
\raisebox{-13.0pt}{\drawsquare{10.0}{0.4}}\hskip-10.4pt
\raisebox{-3.0pt}{\drawsquare{10.0}{0.4}}\hskip-10.4pt
        \raisebox{7pt}{\drawsquare{10.0}{0.4}}\hskip-0.4pt
\raisebox{-3.0pt}{\drawsquare{10.0}{0.4}}\hskip-10.4pt
        \raisebox{7pt}{\drawsquare{10.0}{0.4}}\hskip-0.4pt
\raisebox{-3.0pt}{\drawsquare{10.0}{0.4}}\hskip-10.4pt
        \raisebox{7pt}{\drawsquare{10.0}{0.4}}\hskip-0.4pt
\cdots
\raisebox{-3.0pt}{\drawsquare{10.0}{0.4}}\hskip-10.4pt
        \raisebox{7pt}{\drawsquare{10.0}{0.4}}\hskip-0.4pt\,,
S=\frac12\right) +
\left(\,\,
\raisebox{-8.0pt}{\drawsquare{10.0}{0.4}}\hskip-10.4pt
        \raisebox{2pt}{\drawsquare{10.0}{0.4}}\hskip-0.4pt
\raisebox{-8.0pt}{\drawsquare{10.0}{0.4}}\hskip-10.4pt
        \raisebox{2pt}{\drawsquare{10.0}{0.4}}\hskip-0.4pt
\cdots
\raisebox{-8.0pt}{\drawsquare{10.0}{0.4}}\hskip-10.4pt
        \raisebox{2pt}{\drawsquare{10.0}{0.4}}\hskip-0.4pt
        \raisebox{2pt}{\drawsquare{10.0}{0.4}}\hskip-0.4pt
        \raisebox{2pt}{\drawsquare{10.0}{0.4}}\hskip-0.4pt
        \raisebox{2pt}{\drawsquare{10.0}{0.4}}\hskip-0.4pt\,,
S=\frac12,\frac32,\frac52\right)\\
&+&
\left(\,\,
\raisebox{-8.0pt}{\drawsquare{10.0}{0.4}}\hskip-10.4pt
        \raisebox{2pt}{\drawsquare{10.0}{0.4}}\hskip-0.4pt
\raisebox{-8.0pt}{\drawsquare{10.0}{0.4}}\hskip-10.4pt
        \raisebox{2pt}{\drawsquare{10.0}{0.4}}\hskip-0.4pt
\cdots
\raisebox{-8.0pt}{\drawsquare{10.0}{0.4}}\hskip-10.4pt
        \raisebox{2pt}{\drawsquare{10.0}{0.4}}\hskip-0.4pt
        \raisebox{2pt}{\drawsquare{10.0}{0.4}}\hskip-0.4pt\,,
S=\frac12,\frac32\right) +
\left(\,\,
\raisebox{-13.0pt}{\drawsquare{10.0}{0.4}}\hskip-10.4pt
\raisebox{-3.0pt}{\drawsquare{10.0}{0.4}}\hskip-10.4pt
        \raisebox{7pt}{\drawsquare{10.0}{0.4}}\hskip-0.4pt
\raisebox{-3.0pt}{\drawsquare{10.0}{0.4}}\hskip-10.4pt
        \raisebox{7pt}{\drawsquare{10.0}{0.4}}\hskip-0.4pt
\raisebox{-3.0pt}{\drawsquare{10.0}{0.4}}\hskip-10.4pt
        \raisebox{7pt}{\drawsquare{10.0}{0.4}}\hskip-0.4pt
\cdots
\raisebox{-3.0pt}{\drawsquare{10.0}{0.4}}\hskip-10.4pt
        \raisebox{7pt}{\drawsquare{10.0}{0.4}}\hskip-0.4pt
        \raisebox{7pt}{\drawsquare{10.0}{0.4}}\hskip-0.4pt
        \raisebox{7pt}{\drawsquare{10.0}{0.4}}\hskip-0.4pt\,,
S=\frac12,\frac32\right)\nonumber\\
&+&\cdots + \left(\,\,
\overbrace{\,
\raisebox{-3pt}{\drawsquare{10.0}{0.4}}\hskip-0.4pt
\raisebox{-3pt}{\drawsquare{10.0}{0.4}}\hskip-0.4pt\,\cdots\,
\raisebox{-3pt}{\drawsquare{10.0}{0.4}}\hskip-0.4pt\,}^{N_c}\,,
S=\frac{N_c}{2}-1\right) +
\left(\,\,
\overbrace{\,
\raisebox{-8.0pt}{\drawsquare{10.0}{0.4}}\hskip-10.4pt
        \raisebox{2pt}{\drawsquare{10.0}{0.4}}\hskip-0.4pt
\raisebox{2pt}{\drawsquare{10.0}{0.4}}\,\cdots\,
\raisebox{2pt}{\drawsquare{10.0}{0.4}}\,}^{N_c-1}\,,
S=\frac{N_c}{2}-1,\frac{N_c}{2}\right)\,.\nonumber
\eea

The physical multiplets with well-defined spin $J$ are obtained by adding
the orbital angular momentum $\vec J=\vec S+\vec L$ with $L=1$.

The first three SU(3) representations on the right-hand
side of (\ref{SU(6)mixed}) correspond for $N_c=3$ to {\bf 1}, {\bf 10} 
and {\bf 8} respectively. The others are new and appear only for 
$N_c>3$. Their isospin content for each value of the strangeness number
$K=n_s/2$ can be read off from the corresponding weight diagrams and
is given below for the first few representations.

\bea\label{singlet}
& &
\raisebox{-13.0pt}{\drawsquare{10.0}{0.4}}\hskip-10.4pt
\raisebox{-3.0pt}{\drawsquare{10.0}{0.4}}\hskip-10.4pt
        \raisebox{7pt}{\drawsquare{10.0}{0.4}}\hskip-0.4pt
%\hskip-0.4pt
\raisebox{-3.0pt}{\drawsquare{10.0}{0.4}}\hskip-10.4pt
        \raisebox{7pt}{\drawsquare{10.0}{0.4}}\hskip-0.4pt
\raisebox{-3.0pt}{\drawsquare{10.0}{0.4}}\hskip-10.4pt
        \raisebox{7pt}{\drawsquare{10.0}{0.4}}\hskip-0.4pt
\raisebox{-3.0pt}{\drawsquare{10.0}{0.4}}\hskip-10.4pt
        \raisebox{7pt}{\drawsquare{10.0}{0.4}}\hskip-0.4pt \to
 (K=\frac12\,, I=0) + (K=1\,, I=\frac12) + \cdots\\
& &\label{decuplet}
\raisebox{-8.0pt}{\drawsquare{10.0}{0.4}}\hskip-10.4pt
        \raisebox{2pt}{\drawsquare{10.0}{0.4}}\hskip-0.4pt
\raisebox{-8.0pt}{\drawsquare{10.0}{0.4}}\hskip-10.4pt
        \raisebox{2pt}{\drawsquare{10.0}{0.4}}\hskip-0.4pt
\raisebox{-8.0pt}{\drawsquare{10.0}{0.4}}\hskip-10.4pt
        \raisebox{2pt}{\drawsquare{10.0}{0.4}}\hskip-0.4pt
        \raisebox{2pt}{\drawsquare{10.0}{0.4}}\hskip-0.4pt
        \raisebox{2pt}{\drawsquare{10.0}{0.4}}\hskip-0.4pt
        \raisebox{2pt}{\drawsquare{10.0}{0.4}}\hskip-0.4pt \to
(K=0\,, I=\frac32) + (K=\frac12\,, I=1,2) + (K=1\,, I=\frac12,\frac32,\frac52)
 + \cdots\\
& &\label{octet}
\raisebox{-8.0pt}{\drawsquare{10.0}{0.4}}\hskip-10.4pt
        \raisebox{2pt}{\drawsquare{10.0}{0.4}}\hskip-0.4pt
\raisebox{-8.0pt}{\drawsquare{10.0}{0.4}}\hskip-10.4pt
        \raisebox{2pt}{\drawsquare{10.0}{0.4}}\hskip-0.4pt
\raisebox{-8.0pt}{\drawsquare{10.0}{0.4}}\hskip-10.4pt
        \raisebox{2pt}{\drawsquare{10.0}{0.4}}\hskip-0.4pt
        \raisebox{2pt}{\drawsquare{10.0}{0.4}}\hskip-0.4pt \to
(K=0\,, I=\frac12) + (K=\frac12\,, I=0,1) + (K=1\,, I=\frac12,\frac32)
 + \cdots\\
& &\label{new}
\raisebox{-13.0pt}{\drawsquare{10.0}{0.4}}\hskip-10.4pt
\raisebox{-3.0pt}{\drawsquare{10.0}{0.4}}\hskip-10.4pt
        \raisebox{7pt}{\drawsquare{10.0}{0.4}}\hskip-0.4pt
\raisebox{-3.0pt}{\drawsquare{10.0}{0.4}}\hskip-10.4pt
        \raisebox{7pt}{\drawsquare{10.0}{0.4}}\hskip-0.4pt
\raisebox{-3.0pt}{\drawsquare{10.0}{0.4}}\hskip-10.4pt
        \raisebox{7pt}{\drawsquare{10.0}{0.4}}\hskip-0.4pt
\raisebox{-3.0pt}{\drawsquare{10.0}{0.4}}\hskip-10.4pt
        \raisebox{7pt}{\drawsquare{10.0}{0.4}}\hskip-0.4pt
        \raisebox{7pt}{\drawsquare{10.0}{0.4}}\hskip-0.4pt
        \raisebox{7pt}{\drawsquare{10.0}{0.4}}\hskip-0.4pt \to
(K=\frac12\,, I=1) + (K=1\,, I=\frac12,\frac32)
 + \cdots\,.
\eea
All the other SU(3) multiplets in (\ref{SU(6)mixed}) contain,
for $K=0$, isospin multiplets with $I \geq 2$.
Let us consider now in turn the sectors with different values 
of the strangeness number $K=n_s/2$.

\subsection{K=0}

We list in Table 1 the lowest-lying $K=0$ p-wave light baryons 
containing only $u,d$ quarks. They are contained in the SU(3) 
representations (\ref{singlet},\ref{decuplet},\ref{octet}) which
will be called in the following {\bf 1}, {\bf 10} and {\bf 8}
respectively, corresponding to their dimension for $N_c=3$.

\begin{center}
\begin{tabular}{|c|c|c|c|c|}
\hline
State & \quad $(I,J^P)$\quad\quad & \quad $\Delta$\quad\quad &
\quad $(I,S)$\quad\quad & $(SU(3),SU(2))$ \\
\hline
\hline
N(1535) & $(\frac12,\frac12^-)$ & 1 & $(\frac12,\frac12)$ & 
$({\bf 8}, {\bf 2})$ \\
N(1520) & $(\frac12,\frac32^-)$ &  & &  \\
\hline
N(1650) & $(\frac12,\frac12^-)$ & 0 & & \\
\cline{1-3}
N(1700) & $(\frac12,\frac32^-)$ & 2 & $(\frac12,\frac32)$ &
 $({\bf 8}, {\bf 4})$ \\
N(1675) & $(\frac12,\frac52^-)$ &  &  &  \\
\hline
$\Delta(1620)$ & $(\frac32,\frac12^-)$ & -- & $(\frac32,\frac12)$ & 
$({\bf 10}, {\bf 2})$ \\
$\Delta(1700)$ & $(\frac32,\frac32^-)$ & -- &  &  \\
\hline
\end{tabular}
\end{center}
\begin{quote}
{\bf Table 1.}
The p-wave light baryons containing only $u,d$ quarks and their quantum 
numbers.
\end{quote}

The entries in the last three columns of this table require some explanation.
Usually these states are labeled by the quark model quantum numbers $(I,S)$, 
the total isospin and spin of the quarks. The assignments shown in Table 1 
for this quantum number are the conventional ones \cite{PDG}. Of course, in 
Nature $S$ is not 
a good quantum numbers and the physical eigenstates of $(I,J)$ are linear 
combinations of states with different values of $S$. This mixing is usually
considered to have a dynamical origin and is treated in a phenomenological way.

The large-$N_c$ treatment of these states discussed in \cite{1} suggests a 
different
picture. In this approach the physical states are classified into
towers of states, each labelled by a spin vector $\Delta$. The members of 
a given
tower have quantum numbers $(I,J)$ which are constrained by the condition
$|I-J|\leq\Delta$ and are degenerate in the large-$N_c$ limit. $1/N_c$ 
corrections 
will in general remove this degeneracy and will split the states of the tower.

The connection between the tower states and the $(I,S)$ quark model states has 
been given in \cite{1} for states containing only $u$ and $d$ quarks 
(Eq.(3.23) in \cite{1}) 
\bea\label{3Jrecoupling}
& &|I,(PL)\Delta;Jm\alpha\rangle =\\
& &\qquad\qquad (-)^{I+P+L+J}\sum_S
\sqrt{(2S+1)(2\Delta+1)}
\left\{ \begin{array}{ccc}
I & P & S \\
L & J & \Delta \end{array}\right\}
|(IP)S,L;Jm\alpha\rangle\,.\nonumber
\eea
Here $P=1$ is the so-called $P$-spin introduced in \cite{1} to relate $I$ and
$S$ for quark model states transforming under the mixed symmetry representation
of SU(4). For the p-wave states in Table 1 one has $L=1$. 
One can see that in general the tower states do not have well-defined values
of $S$ and the relation (\ref{3Jrecoupling}) yields the following mixing 
matrices.

The sector $(I,J)=(\frac12,\frac12)$.
\bea\label{mix1}
& &|I=\frac12,\Delta=0;J=\frac12 \rangle =
-\frac{1}{\sqrt3} |I=\frac12,S=\frac12;J=\frac12 \rangle +
\sqrt{\frac23} |I=\frac12,S=\frac32;J=\frac12 \rangle\\
& &|I=\frac12,\Delta=1;J=\frac12 \rangle =
\sqrt{\frac23} |I=\frac12,S=\frac12;J=\frac12 \rangle +
\frac{1}{\sqrt3} |I=\frac12,S=\frac32;J=\frac12 \rangle
\eea

The sector $(I,J)=(\frac12,\frac32)$.
\bea
& &|I=\frac12,\Delta=1;J=\frac32 \rangle =
-\frac{1}{\sqrt6} |I=\frac12,S=\frac12;J=\frac32 \rangle +
\sqrt{\frac56} |I=\frac12,S=\frac32;J=\frac32 \rangle\\\label{mix4}
& &|I=\frac12,\Delta=2;J=\frac32 \rangle =
\sqrt{\frac56} |I=\frac12,S=\frac12;J=\frac32 \rangle +
\frac{1}{\sqrt6} |I=\frac12,S=\frac32;J=\frac32 \rangle
\eea

An examination of the mass spectrum of the $I=\frac12$ states in
Table 1 suggests their association into towers of states with the
shown values of $\Delta$. The relations (\ref{mix1}-\ref{mix4}) give 
then a prediction for the mixing matrices of these states, which can 
be compared with experimental data. Adopting the definitions of 
\cite{CGKM} the mixing of the $N$ states is parametrized as
\bea
N(1650) &=& \cos\theta_{N_1} |S=\frac32\rangle - \sin\theta_{N_1} 
|S=\frac12\rangle\\
N(1535) &=& \cos\theta_{N_1} |S=\frac12\rangle + \sin\theta_{N_1} 
|S=\frac32\rangle
\eea
and 
\bea
N(1520) &=& \cos\theta_{N_3} |S=\frac12\rangle + \sin\theta_{N_3} 
|S=\frac32\rangle\\
N(1700) &=& -\sin\theta_{N_3} |S=\frac12\rangle + \cos\theta_{N_3} 
|S=\frac32\rangle\,.
\eea
We obtain from (\ref{mix1}-\ref{mix4}) the following predictions for the
mixing angles $\theta_{N_1}=0.615\,,\theta_{N_3}=1.991$. The fit of 
\cite{CGKM} to
the strong decays of these states gave the results $\theta_{N_1}=0.61\pm 0.09$
and $(\theta_{N_3})_{fit\, 1}=3.04\pm 0.15$, $(\theta_{N_3})_{fit\, 2}=
2.60\pm 0.16$.
The result for $\theta_{N_1}$ is in excellent agreement with the data. 
The disagreement on $\theta_{N_3}$ can probably be ascribed to finite-$N_c$
corrections. Indeed, due to the fictitious nature of the $P$-spin (which 
becomes apparent in the fact that the states $S=I=N_c/2$ are forbidden), 
one expects the deviations from the large-$N_c$ mixing (\ref{3Jrecoupling}) 
to be largest for $S$, $I$ approaching their maximal values $N_c/2$.

The $I=1/2$ states of the towers belong to SU(3) ``octets'' whose Young diagram 
is shown in (\ref{octet}). There are five such octets, two with $J=1/2$, two
with $J=3/2$ and one with $J=5/2$. The large-$N_c$ mass spectrum of the $K=0$
towers constrains therefore the mass spectrum of these octets, which are predicted
to be degenerate in pairs with $J=(1/2,3/2)$ and $J=(3/2,5/2)$, corresponding to
$\Delta=1$ and 2 in the $K=0$ sector respectively. This is very different from the
picture suggested by the quark model, where one expects these octets to fall
into two groups with $J=(1/2,3/2)$ and $J=(1/2,3/2,5/2)$, corresponding to the
two values taken by the total quark spin $S=1/2,3/2$. One problem with the quark
model picture is the inversion of the two levels with $J=3/2$ and $J=5/2$, 
which is difficult to understand by assuming a spin-orbit interaction alone 
\cite{Close,IsgKa}.

The mixing of the octets with identical values of $J$ can be predicted from the
mixings in the $K=0$ sector (\ref{mix1}-\ref{mix4}). These relations
can be extended to all the states in these multiplets as
\bea\label{8mix1}
& &|{\bf 8},J=\frac12 \rangle_{\Delta=0} =
-\frac{1}{\sqrt3} |{\bf 8},J=\frac12 \rangle_{S=1/2} +
\sqrt{\frac23} |{\bf 8},J=\frac12 \rangle_{S=3/2}\\
& &|{\bf 8},J=\frac12 \rangle_{\Delta=1} =
\sqrt{\frac23} |{\bf 8},J=\frac12 \rangle_{S=1/2} +
\frac{1}{\sqrt3} |{\bf 8},J=\frac12 \rangle_{S=3/2}
\eea
and
\bea
& &|{\bf 8},J=\frac32 \rangle_{\Delta=1} =
-\frac{1}{\sqrt6} |{\bf 8},J=\frac32 \rangle_{S=1/2} +
\sqrt{\frac56} |{\bf 8},J=\frac32 \rangle_{S=3/2}\\\label{8mix4}
& &|{\bf 8},J=\frac32 \rangle_{\Delta=2} =
\sqrt{\frac56} |{\bf 8},J=\frac32 \rangle_{S=1/2} +
\frac{1}{\sqrt6} |{\bf 8},J=\frac32 \rangle_{S=3/2}\,.
\eea
The notation $|{\bf 8},J\rangle_\Delta$ does not imply that all the
states of the {\bf 8} belong to a $\Delta$-tower but only labels
the SU(3) representation in terms of its $K=0$ members.

Unfortunately, no unambiguous tower assignments can be made for the
excited $I=\frac32$ $\Delta$ baryons. Because of the small value of $N_c$ in 
the real world the tower structure for $I=\frac32$ is incomplete. For 
example, instead
of a total number of two states with $(I,J)=(\frac32,\frac12)$ expected in the
large-$N_c$ limit, there is only one such state. To fill up all the
$I=\frac32$ members of the towers with $\Delta=0,1,2$, additional states 
would be
required with $(I,S)= (\frac32,\frac32),(\frac32,\frac52)$, which however do 
not appear for $N_c=3$. This problem did not exist for $s$-wave baryons and
has as consequence an unfortunate loss of predictive power for the large-$N_c$
expansion when applied to the excited baryons.

\subsection{K=1/2}

The observed and expected p-wave baryons with one strange quark are listed
in Table 2, together with their quantum numbers. In the quark model these
states are labelled by $(I,S)$ with $S$ the total spin of the quarks in the
baryon. As discussed above, physical states are in general linear combinations
of quark model states with different values of $S$.
The large-$N_c$ expansion combined with SU(3) symmetry can be used to 
predict this mixing.

From the point of view of large-$N_c$ QCD the observed $K=1/2$ states fall 
into 7 towers of states, three towers with $\Delta=1/2$, three towers with
$\Delta=3/2$ and one tower with $\Delta=5/2$. 
Although the tower structure is complete only for the lowest value of the
isospin $I=0$, we can use SU(3) symmetry to assign the $\Sigma$ states in
the octets well-defined values of $\Delta$. However, just as in the case
of the $I=3/2$ states in the $K=0$ sector, this cannot be done in
an unambiguous way for the decuplet baryons. Therefore we cannot make
predictions for the couplings of these states.

States with the same quantum numbers will mix in the general case.
We parametrize this mixing in the $I=0$ sector as in \cite{CGKM} in terms of
six angles. For the $J=1/2$ states we introduce three angles $\theta_{1i}$
with $i=1,2,3$ as
\bea\label{mixing1/2}
\left(
\begin{array}{c}
\Lambda(1670) \\
\Lambda(1800) \\
\Lambda(1405)
\end{array}\right) =
\left(
\begin{array}{ccc}
c_{11} c_{12} & s_{11} c_{12} & s_{12} \\
-s_{11} c_{13} - c_{11} s_{13} s_{12} & c_{11} c_{13} - s_{11} s_{12} s_{13} &
s_{13} c_{12} \\
s_{11} s_{13} - c_{11} c_{13} s_{12} & -c_{11} s_{13} - s_{11} c_{13} s_{12} &
c_{13} c_{12}
\end{array}\right)
\left(
\begin{array}{c}
\Lambda_{11} \\
\Lambda_{31} \\
\mbox{Singlet}_{11}
\end{array}\right)\,,
\eea
with $c_{11}=\cos\theta_{11}\,, s_{11}=\sin\theta_{11}$, etc.

The quark model states on the RHS are denoted as $\Lambda_{2S,2J}$.
In the SU(3) limit two of the angles vanish $\theta_{12}=\theta_{13}=0$, as
there is no mixing between the singlet and octet.
The third angle $\theta_{11}$ can be determined by noting that
some of the $I=0$ states $\Lambda$ belong to the same SU(3) ``octets''
as the $K=0$ states. 
%This correspondence is shown in Fig.1.
Therefore (\ref{8mix1}-\ref{8mix4}) can be used to
obtain their relation to the quark model states with well-defined $S$
and we find $\theta_{11}=0.615$.

\begin{center}
\begin{tabular}{|c|c|c|c|c|}
\hline
State & \quad $(I,J^P)$\quad\quad & \quad$\Delta$\quad\quad &
\quad $(I,S)$\quad\quad & $(SU(3),SU(2))$ \\
\hline
\hline
$\Lambda(1405)$ & $(0,\frac12^-)$ & $\frac12$ & $(0,\frac12)$ & 
$({\bf 1}, {\bf 2})$ \\
\cline{1-3}
$\Lambda(1520)$ & $(0,\frac32^-)$ & $\frac32$ & &  \\
\hline
$\Lambda(1670)$ & $(0,\frac12^-)$ & $\frac12$ & $(0,\frac12)$ & 
$({\bf 8}, {\bf 2})$ \\
\cline{1-2}
$\Sigma(1620)$ & $(1,\frac12^-)$ & & $(1,\frac12)$ & \\
\cline{1-3}
$\Lambda(1690)$ & $(0,\frac32^-)$ & $\frac32$ & $(0,\frac12)$ &  \\
\cline{1-2}
$\Sigma(1670)$ & $(1,\frac32^-)$ &  & $(1,\frac12)$ & \\
\hline
$\Lambda(1800)$ & $(0,\frac12^-)$ & $\frac12$ & $(0,\frac32)$ & 
$({\bf 8}, {\bf 4})$ \\
\cline{1-2}
$\Sigma(1750)$ & $(1,\frac12^-)$ &   & $(1,\frac32)$ & \\
\cline{1-3}
$\Lambda(?)$ & $(0,\frac32^-)$ & $\frac32$ & $(0,\frac32)$ &  \\
\cline{1-2}
$\Sigma(?)$ & $(1,\frac32^-)$ &  & $(1,\frac32)$ & \\
\cline{1-3}
$\Lambda(1830)$ & $(0,\frac52^-)$ & $\frac52$ & $(0,\frac32)$ &  \\
\cline{1-2}
$\Sigma(1775)$ & $(1,\frac52^-)$ &  & $(1,\frac32)$ & \\
\hline
$\Sigma(?)$ & $(1,\frac12^-)$ & -- & $(1,\frac12)$ & 
$({\bf 10}, {\bf 2})$ \\
$\Sigma(?)$ & $(1,\frac32^-)$ & -- & & \\
\hline
\end{tabular}
\end{center}
\begin{quote}
{\bf Table 2.}
The p-wave hyperons containing one strange quark and their quantum 
numbers. $(I,S)$ denote the usual quark model assignments of the
states and $\Delta$ gives their large-$N_c$ tower assignment.
\end{quote}

The sector $J=3/2$ can be treated in an analogous way. The mixing of these
states is parametrized in terms of three angles $\theta_{3i}$ defined as
\bea
\left(
\begin{array}{c}
\Lambda(1690) \\
\Lambda(?) \\
\Lambda(1520)
\end{array}\right) =
R(\theta_{31},\theta_{32},\theta_{33})
\left(
\begin{array}{c}
\Lambda_{13} \\
\Lambda_{33} \\
\mbox{Singlet}_{13}
\end{array}\right)
\eea
where the unitary matrix $R$ is defined in analogy to the one in (\ref{mixing1/2}).
We find for this case, in the limit of SU(3) symmetry, $\theta_{31}=1.991\,,
\theta_{32}=\theta_{33}=0$. Similar predictions can be made in the limit of
SU(3) symmetry for the mixing matrix of the $\Sigma$ states.

The experimental situation with these angles is not very clear. The fit of \cite{CGKM}
gave six different possible solutions for the $\theta_{1i}$ and four solutions
for $\theta_{3i}$. The values taken by the angles $\theta_{i2}$, $\theta_{i3}$
in these solutions do not come close to the SU(3) value (0), which can
be explained by a sizable violation of SU(3) symmetry. This implies in turn the 
existence of similar large deviations from the SU(3)-based prediction for $\theta_{i1}$.
However, the large-$N_c$ predictions for decays of tower states to be presented in
the next Section do not depend on a precise knowledge of the mixing matrix.

\section{Strong Decays}

Let us first recapitulate the results obtained in \cite{1} for strong decays
of excited baryons in the large-$N_c$ limit by assuming only isospin symmetry.
Excited baryons can decay to s-wave baryons through pion emission in
$S$-wave and $D$-wave. The respective couplings are related to matrix
elements of the axial current taken between tower states $(\Delta\to \Delta')$
\bea\label{Ydef}
\langle J'I';m',\alpha' |\bar q\gamma^0\gamma_5\frac12 \tau^aq|JI;m,\alpha\rangle
&=& N_c^\kappa  \langle J'I';m',\alpha' |Y^{a}|JI;m,\alpha\rangle\\\label{Qdef}
\langle J'I';m',\alpha' |\bar q\gamma^i\gamma_5\frac12 \tau^aq|JI;m,\alpha\rangle
&=& N_c^\kappa  q^j
\langle J'I';m',\alpha' |Q^{ij,a}|JI;m,\alpha\rangle\\
&+& N_c^\kappa  \epsilon_{ijk}q^j
\langle J'I';m',\alpha' |R^{k,a}|JI;m,\alpha\rangle\nonumber
\eea
with $q^\mu$ the momentum of the current. 
$\kappa$=0 for a decaying state transforming under the mixed symmetry representation
of SU(4).
The operators $Y^a$ and $Q^{ij,a}$ 
parametrize the $S$-wave and $D$-wave pion couplings respectively.
Their matrix elements are determined, at leading order in $N_c$, by 
four reduced matrix elements $c(\Delta',\Delta),c_{1-3}(\Delta',\Delta)$
\bea\label{SU(2)S}
\langle J'I';m',\alpha' |Y^{a}|JI;m,\alpha\rangle &=&
c(\Delta',\Delta)\sqrt{2I+1}(-)^{I-J-\Delta'}\delta_{JJ'}\delta_{mm'}
\left\{
\begin{array}{ccc}
I' & 1 & I \\
\Delta & J & \Delta'
\end{array}\right\}
\langle I'\alpha'|I1;\alpha a\rangle\\\label{SU(2)D}
\langle J'I';m',\alpha' |Q^{ka}|JI;m,\alpha\rangle &=&
(-)^{J+I+J'+I'}\sqrt{(2J+1)(2I+1)}\\
& &\times \sum_{y=1,2,3}
c_y(\Delta',\Delta)
\left\{
\begin{array}{ccc}
\Delta' & I' & J' \\
\Delta & I & J \\
y & 1 & 2
\end{array}\right\}
\langle J'm'|J2;m k\rangle
\langle I'\alpha'|I1;\alpha a\rangle\nonumber\,.
\eea

In the following we will extend these results to the case of SU(3) symmetry.
%The SU(3) multiplet structure of the lowest-lying excited baryons is 
%shown in Fig.1. 
As explained above, we will restrict our considerations to octet and
singlet states. There are five octets and two singlets, which will be 
represented by SU(3) tensors constructed as in \cite{DJM1}. 

The spin-1/2 octet whose $K=0$ members belong to the $\Delta=0$ tower will
be represented by the tensor $({\cal B}_1)^i_{j_1 j_2 \cdots j_\nu}$ with 
one upper and $\nu=(N_c-1)/2$ lower indices.
The two spin-1/2 and 3/2 octets whose $K=0$ members belong to the $\Delta=1$ 
tower are represented by the tensors $({\cal B}_2)^i_{j_1 j_2 \cdots j_\nu}$ 
and $({\cal B}_3)^i_{j_1 j_2 \cdots j_\nu}$ respectively.
Finally, the two spin-3/2 and 5/2 octets whose $K=0$ members belong to the 
$\Delta=2$ tower will be assigned the tensors $({\cal B}_4)^i_{j_1 j_2 \cdots 
j_\nu}$ and $({\cal B}_5)^i_{j_1 j_2 \cdots j_\nu}$ respectively.

The spin-1/2 and 3/2 singlet baryons are each represented by a SU(3) tensor 
with $\nu-1$ lower indices $({\cal S}_1)_{j_1 j_2 \cdots j_{\nu-1}}$ and
$({\cal S}_2)_{j_1 j_2 \cdots j_{\nu-1}}$.
The nonvanishing components of these tensors for the $\Lambda$ states are 
${\cal S}_{33 \cdots 3}=1$. For $N_c = 3$ these tensors go over into SU(3) 
scalars, as they should.

The s-wave baryons are represented by the usual octet tensor 
${\cal B}^i_{j_1 j_2 \cdots j_\nu}$ (for the spin-1/2 baryons) and the decuplet 
tensor ${\cal T}^{i_1 i_2 i_3}_{j_1 j_2 \cdots j_{\nu-1}}$ (for the spin-3/2 
baryons).

The couplings of the Goldstone bosons are described by interaction Lagrangians
built out of the SU(3) tensors introduced above. The part containing the $S$-wave 
couplings is written in terms of seven SU(3) invariants ${\cal M}_{1,2}\,,
{\cal N}_{1,2}\,,{\cal L}_{1,2}\,,{\cal P}_1$ as
\bea\label{SU(3)S}
{\cal L}_S &=& 
{\cal M}_1 \mbox{tr }(\bar {\cal B}\gamma_\mu A^\mu {\cal B}_1) +
{\cal N}_1 \mbox{tr }(\bar {\cal B}\gamma_\mu {\cal B}_1 A^\mu)\\
&+&
{\cal M}_2 \mbox{tr }(\bar {\cal B}\gamma_\mu A^\mu {\cal B}_2) +
{\cal N}_2 \mbox{tr }(\bar {\cal B}\gamma_\mu {\cal B}_2 A^\mu)\nonumber\\
&+& 
{\cal L}_1 \mbox{tr } (\bar {\cal T}_\mu \gamma_\nu A^\nu {\cal B}_3^\mu)
+ {\cal L}_2 \mbox{tr } (\bar {\cal T}_\mu \gamma_\nu A^\nu
{\cal B}_4^\mu)\nonumber\\
&+& {\cal P}_1 \mbox{tr } (\bar {\cal B}\gamma_\mu A^\mu {\cal S}_1)\nonumber\,.
\eea
The nonlinear axial current field $A_\mu$ is defined by 
$A_\mu=i/2(\xi^\dagger \partial_\mu\xi - \xi\partial_\mu \xi^\dagger)$
with $\xi=\exp(iM/f_\pi)$ and $f_\pi=132$ MeV. The matrix $M$ contains the 
Goldstone boson fields and is given by $M=\frac{1}{\sqrt2}\pi^a \lambda^a$.

The $D$-wave couplings of the Goldstone bosons are described by an analogous
Lagrangian containing twelve SU(3) invariants
\bea\label{SU(3)D}
{\cal L}_D &=& 
m_B {\cal M}_3 \mbox{tr }(\bar {\cal B} A^\mu \gamma_5 {\cal B}_3^\mu) +
m_B {\cal N}_3 \mbox{tr }(\bar {\cal B}\gamma_5 {\cal B}_3^\mu A_\mu)\\
&+&
m_B {\cal M}_4 \mbox{tr }(\bar {\cal B} A^\mu \gamma_5 {\cal B}_4^\mu) +
m_B {\cal N}_4 \mbox{tr }(\bar {\cal B}\gamma_5 {\cal B}_4^\mu A_\mu)
\nonumber\\
&+&
{\cal M}_5 \mbox{tr }(\bar {\cal B} (D_\mu A_\nu+D_\nu A_\mu) {\cal B}_5^{\mu\nu}) +
{\cal N}_5 \mbox{tr }(\bar {\cal B} {\cal B}_5^{\mu\nu}(D_\mu A_\nu+D_\nu A_\mu))
\nonumber\\
&+&
m_T {\cal L}_3 \mbox{tr } (\bar {\cal T}_\mu A^\mu\gamma_5 {\cal B}_1)
+ m_T {\cal L}_4 \mbox{tr } (\bar {\cal T}_\mu A^\mu \gamma_5 {\cal B}_2)
\nonumber\\
&+& 
i{\cal L}_5 \mbox{tr } (\bar {\cal T}^\mu (D_\mu A_\nu + D_\nu A_\mu + \mbox{ t.t.})
{\cal B}_3^\nu) +
i{\cal L}_6 \mbox{tr } (\bar {\cal T}^\mu (D_\mu A_\nu + D_\nu A_\mu + \mbox{ t.t.})
{\cal B}_4^\nu)\nonumber\\
&+& {\cal L}_7 i\varepsilon_{\alpha\beta\gamma\delta}\mbox{tr } 
(\bar {\cal T}^\alpha (D^\rho A^\beta + D^\beta A^\rho) v^\gamma B_5^{\delta\rho})
+ m_B {\cal P}_2 \mbox{tr } (\bar {\cal B}\gamma_5 A_\mu {\cal S}_2^\mu)
\nonumber\,.
\eea
We extracted factors of
$m_B, m_T$ in the definition of some couplings such that their expansion
in powers of $1/N_c$ starts with a term of O(1). The form of the trace terms 
``t.t.'', needed to project out a pure $D$-wave, is given in the Appendix.
In these expressions only the Lorentz indices are written explicitly. 
The traces over the SU(3) indices have the following form: 

a) octet-octet coupling 
\bea\nonumber
\mbox{tr }(\bar {\cal B} A {\cal B}_1) =
\bar {\cal B}^{b_1 b_2 \cdots b_\nu}_a A^a_c
({\cal B}_1)^c_{b_1 b_2 \cdots b_\nu}\,,\qquad
\mbox{tr }(\bar {\cal B} {\cal B}_1 A) =
\bar {\cal B}^{c b_2 \cdots b_\nu}_a 
({\cal B}_1)^a_{d b_2 \cdots b_\nu} A^d_c
\eea

b) octet-decuplet coupling

\bea\nonumber
\mbox{tr } (\bar {\cal T} A {\cal B}_1) =
\varepsilon^{\alpha\beta\gamma}
\bar {\cal T}_{\alpha\mu\nu}^{b_1 b_2 \cdots b_{\nu-1}}
A^\mu_\beta ({\cal B}_1)^\nu_{\gamma b_1 b_2 \cdots b_{\nu-1}}
\eea

c) octet-singlet coupling

\bea\nonumber
\mbox{tr } (\bar {\cal B} A {\cal S}) =
\bar {\cal B}^{b_1 b_2 \cdots b_\nu}_a A^a_{b_1}
{\cal S}_{b_2 \cdots b_\nu}\,.
\eea

The interplay of the large-$N_c$ predictions (\ref{SU(2)S},\ref{SU(2)D}) with
the SU(3) symmetry leads to significant simplifications in the structure of 
the Lagrangian (\ref{SU(3)S},\ref{SU(3)D}). Thus, the $S$-wave pion couplings 
of the excited baryon octets to ground state baryons are described in this 
limit by just one common reduced matrix element (instead of five, assuming 
only isospin invariance) and in the $D$-wave sector only two independent couplings
are required (instead of seven).

These additional relations can be derived by writing representative transition 
amplitudes in two alternative ways, using the SU(3) and SU(2) relations respectively.
We obtain in this way the following model-independent predictions for the 
$S$-wave couplings
\bea\label{Sa}
{\cal M}_1 &=& {\cal O}(1/N_c)\\\label{Sb}
\frac{{\cal M}_2}{{\cal L}_1} &=& -\frac{2}{\sqrt3}+{\cal O}(1/N_c)\\
{\cal L}_2 &=& {\cal O}(1/N_c)\label{Sc}
\eea
and for the $D$-wave couplings
\bea\label{Da}
{\cal L}_3 &=& {\cal O}(1/N_c)\\\label{Db}
\frac{{\cal M}_3}{{\cal L}_5} &=& -\frac83+{\cal O}(1/N_c)\\\label{Dc}
\frac{{\cal M}_3}{{\cal L}_4} &=& -\frac{2}{\sqrt3}+{\cal O}(1/N_c)\\\label{Dd}
\frac{{\cal M}_4}{{\cal L}_6} &=& -\frac43+{\cal O}(1/N_c)\\\label{De}
\frac{{\cal M}_4}{{\cal L}_7} &=& 4\sqrt{\frac25}+{\cal O}(1/N_c)\\\label{Df}
\frac{{\cal M}_5}{{\cal L}_7} &=& -\frac23+{\cal O}(1/N_c)\,.
\eea
The ${\cal N}$ parameters in the Lagrangians (\ref{SU(3)S},\ref{SU(3)D}) 
contribute to the pion couplings only to subleading order (although they 
contribute to the same order as ${\cal M}$ to the kaon couplings). 
Therefore, in order to obtain information about them, knowledge of the pion 
couplings to next-to-leading order in $1/N_c$ is required. This will 
have to be obtained from model calculations.

In practice the $1/N_c$ corrections to the predictions (\ref{Sa}-\ref{Sc}),
(\ref{Da}-\ref{Df}) can be sizable. In the following we compare these 
predictions against available experimental data on strong decays of these
states. To avoid additional complications related to SU(3) breaking 
effects and a more complex mixing structure, we will restrict ourselves to 
pion decays of nonstrange excited baryons.

 The relation (\ref{Sb}) between $S$-wave amplitudes can be tested by examining 
the ratio of decay widths
\bea
(R_1)_{th} = \frac{\Gamma(N(1535)\to [N\pi])}{\Gamma(N(1520)\to 
[\Delta\pi]_S)} =
5.227\frac{{\cal M}_2^2}{{\cal L}_1^2} = 6.969\,.
\eea
We used on the RHS the theoretical expression for the widths together
with the coupling ratio (\ref{Sb}). The experimental value of this ratio is
\cite{PDG} 
\bea\label{R1exp}
(R_1)_{exp}= 6.625^{+18.35}_{-4.46}\,.
\eea
Not all relations for $S$-wave couplings work as well. For example, one expects
from (\ref{Sa}) the coupling ${\cal M}_1$ to be suppressed by $1/N_c$ relative
to ${\cal M}_2$. However, the corresponding ratio of decay widths
\bea\label{R2}
(R_2)_{th} = \frac{\Gamma(N(1650)\to [N\pi])}{\Gamma(N(1535)\to [N \pi])} =
1.58\frac{{\cal M}_1^2}{{\cal M}_2^2}
\eea
takes the experimental value $(R_2)_{exp} = 0.58-4.88$, which is at least a factor 
of 4 larger than the one obtained with the naive 
estimate ${\cal M}_1^2/{\cal M}_2^2\simeq 0.1$.

The situation with the prediction (\ref{Sc}) is less clear, as the PDG does not
quote branching ratios for the decay mode $N(1700)\to [\Delta\pi]_{S,D}$. The
$S$-wave mode appears however to be suppressed in comparison to the $D$-wave 
one \cite{ManSa}, in agreement with the large-$N_c$ expectation 
from (\ref{Sc}).

This analysis can be extended to the $D$-wave couplings. The following ratios of decay 
widths can be used to test (\ref{Db}), (\ref{Dc}), (\ref{De}) and (\ref{Df}).
\bea\label{R3}
(R_3)_{th} &=& \frac{\Gamma(N(1520)\to [N\pi]_D)}{\Gamma(N(1520)\to [\Delta\pi]_D)} =
2.151\frac{{\cal M}_3^2}{{\cal L}_5^2} = 15.30\,,\quad (R_3)_{exp}=3.57-6.01\\
\label{R4}
(R_4)_{th} &=& \frac{\Gamma(N(1520)\to [N\pi]_D)}{\Gamma(N(1535)\to [\Delta\pi]_D)} =
4.216\frac{{\cal M}_3^2}{{\cal L}_4^2} = 5.62\,,\quad (R_4)_{exp}\geq 4.4\\
\label{R5}
(R_5)_{th} &=& \frac{\Gamma(N(1675)\to [N\pi]_D)}{\Gamma(N(1675)\to [\Delta\pi]_D)} =
4.595\frac{{\cal M}_5^2}{{\cal L}_7^2} = 2.042\,,\quad (R_5)_{exp}=0.66-1.00\\
\label{R6}
(R_6)_{th} &=& \frac{\Gamma(N(1700)\to [N\pi]_D)}{\Gamma(N(1675)\to [\Delta\pi]_D)} =
0.883\frac{{\cal M}_4^2}{{\cal L}_7^2} = 5.651\,,\quad (R_6)_{exp}=0.055-0.2\,.
\eea
We do not present a comparison with data for the ratio (\ref{Dd}) because
of the lack of data on $N(1700)\to [\Delta\pi]_D$.

The deviations of these ratios from the large-$N_c$ predictions can be 
understood
%\footnote{With the notable exception of $R_3$ which is shared also
%by other quark model calculations \cite{CGKM}.}
partly as a consequence of the finite value of $N_c$ and partly because of the 
sensitivity of these ratios to the precise value of the mixing angle 
$\theta_{N_3}$. 
We will use in the following the quark model with $N_c=3$ to illustrate the 
importance
of the $1/N_c$ corrections. The couplings of the $N^*$ states are
related in the quark model to the reduced matrix elements ${\cal T}(I',SI)$
introduced in \cite{1}. Their explicit formulas for arbitrary $N_c$ are 
(normalized to (3.50) of \cite{1} in the large-$N_c$ limit)
\bea
& &{\cal T}(\frac12,\frac12 \frac12) = -\frac{2\sqrt2}{3}
\sqrt{\frac{(N_c-1)(N_c+3)}{N_c(N_c+2)}}{\cal I}\,,\qquad
{\cal T}(\frac12,\frac32 \frac12) = -\frac23
\sqrt{\frac{N_c-1}{N_c+2}}{\cal I}\,,\\
& &{\cal T}(\frac32,\frac12 \frac12) = \frac23
\sqrt{\frac{(N_c+3)(N_c+5)}{N_c(N_c+2)}}{\cal I}\,,\qquad\qquad
{\cal T}(\frac32,\frac32 \frac12) = -\frac23\sqrt5
\sqrt{\frac{N_c+5}{N_c+2}}{\cal I}\nonumber
\eea
with ${\cal I}$ a common overlap integral.
We obtain for example for the ratio (\ref{Sb}) of the $S$-wave couplings
\bea
\frac{{\cal M}_2}{{\cal L}_1} &=& -2\sqrt{\frac23}
\frac{{\cal T}(\frac12,\frac12 \frac12)\cos\theta_{N_1}+{\cal T}(\frac12,\frac32
\frac12)\sin\theta_{N_1}}{{\cal T}(\frac32,\frac12 \frac12)\cos\theta_{N_3}+
{\cal T}(\frac32,\frac32\frac12)\sin\theta_{N_3}}\,.
\eea
In the large-$N_c$ limit and for the mixing angles given in Sec.II.A the value of
this ratio reduces to (\ref{Sb}). For $N_c=3$ it gives \cite{FaiPla,Hey}
\bea\label{SbNc=3}
\frac{{\cal M}_2}{{\cal L}_1} &=& \sqrt{\frac23}
\frac{2\cos\theta_{N_1}+\sin\theta_{N_1}}
{\sqrt2 \cos\theta_{N_3}-\sqrt5\sin\theta_{N_3}}\\
&=& -0.689(\theta_{N_3}=1.991)\,,\quad
-0.763(\theta_{N_3}=2.6)\,,\quad
-1.105(\theta_{N_3}=3.04)\nonumber\,.
\eea
The numerical values shown are computed with the large-$N_c$ value for 
$\theta_{N_1}=0.615$ which was seen to agree well with the experimental one.
For the largest value of $\theta_{N_3}=3.04$, the ratio (\ref{SbNc=3}) 
predicts $(R_1)_{th}=6.382$ which is in good agreement with the experimental
value (\ref{R1exp}).

The ratio ${\cal M}_1/{\cal M}_2$ depends only on the angle $\theta_{N_1}$
and is given by
\bea
\frac{{\cal M}_1}{{\cal M}_2} &=&
\frac{-{\cal T}(\frac12,\frac12 \frac12)\sin\theta_{N_1}+
{\cal T}(\frac12,\frac32\frac12)\cos\theta_{N_1}}
{{\cal T}(\frac12,\frac12 \frac12)\cos\theta_{N_1}+
{\cal T}(\frac12,\frac32\frac12)\sin\theta_{N_1}}\\
& &\to -\frac{2\sin\theta_{N_1}-\cos\theta_{N_1}}
{2\cos\theta_{N_1}+\sin\theta_{N_1}} = (-0.056)-(-0.241)\,,\qquad
(N_c=3)\nonumber\,.
\eea
In the last line we used the experimental value $\theta_{N_1}=0.61\pm 0.09$
\cite{CGKM}. This yields in turn a result for the ratio (\ref{R2}) 
$(R_2)_{th}=0.005-0.092$, which is still smaller than the experimental 
value $(R_2)_{exp}=0.58-4.88$. We will return later to a discussion of 
this discrepancy.

Similar results are obtained for the ratios of $D$-wave couplings. For example,
we get
\bea
\frac{{\cal M}_3}{{\cal L}_5} &=& -\frac43
\frac{2\sqrt5 {\cal T}(\frac12,\frac12 \frac12)\cos\theta_{N_3}-
\sqrt2 {\cal T}(\frac12,\frac32\frac12)\sin\theta_{N_3}}
{\sqrt{\frac52}{\cal T}(\frac32,\frac12 \frac12)\cos\theta_{N_3}-
2\sqrt{\frac25}{\cal T}(\frac32,\frac32\frac12)\sin\theta_{N_3}}\,.
\eea
Taking in this expression $N_c=3$ gives
\bea
\frac{{\cal M}_3}{{\cal L}_5} &=& -\frac43
\frac{-2\sqrt{10}\cos\theta_{N_3}+\sin\theta_{N_3}}
{\sqrt{10}\cos\theta_{N_3}+4\sin\theta_{N_3}}\\
&=& -1.972(\theta_{N_3}=1.991)\,,\quad
12.217(\theta_{N_3}=2.6)\,,\quad
2.493-4.112(\theta_{N_3}=3.04\pm 0.15)\nonumber\,.
\eea
This ratio is particularly sensitive to the mixing angle $\theta_{N_3}$
as the physical value of this angle lies in the vecinity of 2.47, where the
denominator vanishes. The ratio $R_3$ corresponding to $\theta_{N_3}=3.04\pm 0.15$
is still larger by about a factor of 2 than the experimental value (\ref{R3}).
Similar large values for $R_3$ appear to be predicted also in other quark 
model calculations \cite{CGKM}.

The ratio (\ref{Df}) of the couplings of the $J^P=5/2^-$ state is given in the
quark model by
\bea
\frac{{\cal M}_5}{{\cal L}_7} &=&
-\frac23\sqrt5 \frac{{\cal T}(\frac12,\frac32\frac12)}
{{\cal T}(\frac32,\frac32\frac12)} = -\frac23\sqrt{\frac{N_c-1}{N_c+5}}\,.
\eea
For $N_c=3$ this implies $(R_5)_{th}=0.510$ which is in reasonable agreement
(although somewhat smaller) with the experimental result (\ref{R5}).

Finally, the ratio (\ref{De}) is given by
\bea
\frac{{\cal M}_4}{{\cal L}_7} &=& \frac{8}{3\sqrt3}
\frac{\sqrt{10}{\cal T}(\frac12,\frac12\frac12)\sin\theta_{N_3}+
{\cal T}(\frac12,\frac32\frac12)\cos\theta_{N_3}}
{{\cal T}(\frac32,\frac32\frac12)}
\eea
which for $N_c=3$ reduces to
\bea
\frac{{\cal M}_4}{{\cal L}_7} &=&
\frac{4}{3\sqrt{15}}
(\cos\theta_{N_3}+2\sqrt{10}\sin\theta_{N_3})
= 1.847 (\theta_{N_3}=1.991)\,,\\
& &\qquad 0.491-1.142 (\theta_{N_3}=2.6\pm 0.16)\,,\qquad
(-0.449)-(0.209) (\theta_{N_3}=3.04\pm 0.15)\nonumber\,.
\eea
For $\theta_{N_3}=3.04\pm 0.15$ this gives $(R_6)_{th}=0.038-0.178$ which
is in agreement with the experimental value (\ref{R6}).

Perhaps the most puzzling disagreement between the large-$N_c$ predictions
and experiment concerns the large experimental value of the ratio
$R_2$ (\ref{R2}). Among the possible explanations for this disagreement,
we can mention: a) wrong assignments of the $\Delta$ quantum numbers for
the $S_{11}$ states; b) a large deviation of the mixing angle $\theta_{N_1}$
from its predicted value $\theta_{N_1}=0.615$; c) the presence of a third
$S_{11}$ state in the region around 1.6 GeV. The first possibility
entails assigning $\Delta=1$ to $N(1650)$ and $\Delta=0$ to $N(1535)$, which 
results into the prediction $\theta_{N_1}=-0.955$. This would give in turn a
value for the ratio (\ref{R2}) $(R_2)_{th}=67.47$ which is almost a factor
of 5 larger than the one obtained with the dimensional estimate ${\cal M}_1/
{\cal M}_2\simeq N_c=3$. The second alternative b) requires the angle
$\theta_{N_1}$ to be of the order of $-0.08$ or $1.04$. Furthermore, the 
large splitting between the members of the $\Delta=1$ tower in the case a) 
together with the large disagreement
in the value of $\theta_{N_1}$ with other determinations \cite{CGKM}
combine to make these two possible explanations rather unattractive.

Recent analyses of the $\pi N$ scattering data 
\cite{Arndt} show evidence for a new $J^P=1/2^-$ state with a mass of 
1712 MeV.
Since its mass is very close to that of $N(1650)$, it is possible that the
data quoted by the PDG \cite{PDG} referring to the latter in 
fact cummulates over the decays of both states.
It is interesting to note that the new state has a small branching ratio for 
decays into the $N\pi$ mode, of about 20\% \cite{Arndt}, which 
fits the large-$N_c$ prediction for the $\Delta=0$ state.
It is tempting therefore to identify this state with the $J=1/2$ member 
of the $\Delta=0$ tower. It is not yet clear what the quark model 
interpretation of
each of the three $S_{11}$ states is (for example, in \cite{KaiSi} it is 
proposed to interpret one of them as a bound state $\Sigma K$, see also 
\cite{LiWo}). Further investigation of these states is required to help 
settle this apparent puzzle of the large-$N_c$ expansion.

\section*{Conclusions}

We have analyzed in this paper the phenomenological consequences of the
large-$N_c$ expansion for the $L=1$ orbitally excited baryons, following
from the formalism described in \cite{1}. These states are organized into
towers of states, whose couplings to the ground state baryons are related
in a simple way. In the large-$N_c$ limit the members of a given tower
are degenerate, which yields constraints on the masses of these states
which are distinct from those of the quark model with SU(6) symmetry.
Quite remarkably, the mixing angles of the five octets of $L=1$ excited 
baryons are completely predicted in the combined large-$N_c$ and SU(3)
limits. Unfortunately, because of the small value of the $N_c$ parameter
in the real world, we cannot accomodate the decuplet states into the 
picture suggested by large-$N_c$ QCD. Despite these shortcomings, 
we believe that this approach could be used (much in the same way as done 
in \cite{Jen} for the ground state baryons) as the starting point for a 
systematic study of the $1/N_c$ and SU(3) breaking corrections for these 
states.

\acknowledgements
D.P. is grateful for the hospitality extended to him by the Theory Group of 
the National Tsing Hua University, Taiwan and 
to the Center for Theoretical Sciencex for support. His research is also 
supported by the Ministry of Science and the Arts of Israel. The work of 
T.M.Y. was supported in part by the National Science Foundation.

\appendix

\section{}

We present in this Appendix the partial wave decomposition for the decay
$3/2^-\to (3/2^+,0^-)$ which can proceed through both $S$- and $D$-wave. 
The invariant transition matrix element is decomposed as
\bea
{\cal M} &=& \bar u^\mu(p')\left\{
c_D\left( q_\mu q_\nu + 2\vec q\,^2 \frac{m_P^2}{m_P^2+4m_P m_S+m_S^2-q^2}
g_{\mu\nu}\right)\right.\\
& &\left. \qquad\qquad+
c_S\left( g_{\mu\nu} + \frac{2}{(m_S+m_P)^2-q^2}q_\mu q_\nu\right)\right\}
u^\nu(p)\,,\nonumber
\eea
with $\vec q$ the pion 3-momentum in the rest frame of the decaying particle.
The masses of the initial and final particles are denoted as $m_P$ and $m_S$
respectively.
The partial decay widths are given by
\bea
\Gamma_S &=& \frac{1}{8\pi}c_S^2 \frac{(m_S+m_P)^2-q^2}{m_P^2}|\vec q\,|\\
\Gamma_D &=& \frac{1}{2\pi}c_D^2 \frac{m_P^2[(m_S+m_P)^2-q^2]}
  {(m_P^2+4m_P m_S+m_S^2-q^2)^2}|\vec q\,|^5\,.
\eea


\begin{references}
\bibitem{LargeNc} G. t'Hooft, {\em Nucl.Phys.} {\bf B72} 461 (1974);
   {\bf B75} 461 (1974).
\bibitem{Wi} E. Witten, {\em Nucl.Phys.} {\bf B160} 57 (1979).
\bibitem{Coleman} S. Coleman, {\em 1/N} in {\em Aspects of Symmetry: Selected
   Erice Lectures}, Cambridge University Press, Cambridge 1985.
\bibitem{DJM1} R. Dashen, E. Jenkins and A.V. Manohar, {\em Phys.Rev.}
   {\bf D49} 4713 (1994); {\bf D51} 3697 (1995).
\bibitem{DJM2} J. Dai, R. Dashen, E. Jenkins and A.V. Manohar, {\em Phys.Rev.}
   {\bf D53} 273 (1996).
\bibitem{1} D. Pirjol and T.M. Yan, CLNS-97/1500, hep-ph/9707485.
\bibitem{FaiPla} D. Faiman and D. Plane, {\em Nucl.Phys.} {\bf B50} 379
   (1972).
\bibitem{Hey} A. Hey, P. Litchfield and R. Cashmore,
   {\em Nucl.Phys.} {\bf B95} 516 (1975).

\bibitem{Close} F.E. Close, {\em An Introduction to 
   Quarks and Partons}, Academic Press, 1979.
\bibitem{IsgKa} N. Isgur and G. Karl, {\em Phys.Lett.} {\bf B72} 109
   (1977); {\em Phys.Rev.} {\bf D18} 4187 (1978).
\bibitem{ManSa} D.M. Manley and E.M. Saleski, {\em Phys.Rev.}
   {\bf D45} 4002 (1992).
\bibitem{CGKM} C.D. Carone, H. Georgi, L. Kaplan and D. Morin,
   {\em Phys.Rev.} {\bf D50} 5793 (1994).

\bibitem{PDG}  Particle Data Group, R.M. Barnett {\em et al.},
   {\em Phys.Rev.} {\bf D54} 1 (1996).
\bibitem{Arndt} R.A. Arndt, I.I. Strakovsky, R.L. Workman and M.M. Pavan,
   {\em Phys.Rev.} {\bf C52} 2120 (1995)
\bibitem{KaiSi} N. Kaiser, P.B. Siegel and W.Weise,
   {\em Phys.Lett.} {\bf B362} 23 (1995).
\bibitem{LiWo} Z. Li and R. Workman, {\em Phys.Rev.}
   {\bf C53} 549 (1996).
\bibitem{Jen} E. Jenkins, {\em Phys.Rev.} {\bf D53} 2625 (1996).

\end{references}
\end{document}